# Arithmetic with spatiotemporal optical vortex of integer and fractional topological charges

**Hsiao-Chih Huang[a,b], Chen-Ting Liao[b], Hui Min Leung[a,*]**

[a]Department of Intelligent Systems Engineering, Indiana University, Bloomington, IN 47408, USA
[b]Department of Physics, Indiana University, Bloomington, IN 47405, USA

**Abstract**. Spatiotemporal optical vortices carry transverse orbital angular momentum (t-OAM), which give rise to spatiotemporal topological charge (ST-TC). To unleash the full potential of t-OAM in expanding the capacity of communication and computing, we demonstrate the first optical information-processing pipeline capable of performing addition and subtraction on ST-TC values, regardless of whether they are integer or fractional. Additionally, we established a readout method for those mathematical operations through imaging spectral analysis, providing a robust optical basis toward arithmetic operations and verification. These new capabilities mark crucial advancements toward full arithmetic operations on the ST-TC of light for bosonic state computation and information processing.

*Hui Min Leung, E-mail: huileung@iu.edu

## 1 Introduction

Light is a fundamental carrier of information in both classical and quantum communication and computation. Information can be encoded onto different degrees of freedom (DOFs) of light such as polarization, wavelength, amplitude, and phase. The increasing demand for higher bandwidth and high-speed capacity has driven efforts to explore and leverage new DOFs for information encoding and processing. Without the introduction of new optical DOFs, current fiber-optic technologies are fast approaching the data transmission limit due to nonlinear effects in the fibers.

Among various DOFs, optical vortex beams carrying orbital angular momentum (OAM) have gained substantial attention over the past three decades [1]. These OAM beams possess a twisted, helical wavefront characterized by a spatial topological charge (S-TC), which is an integer determined by the number of 2π phase winding in the transverse ($x$–$y$) plane, where the charge can be 0, ±1, ±2, …, ±N, where N is an integer. This unique spatial phase enables a range of applications, including high-resolution imaging [2] and optical metrology and sensing [3, 4]. Due to the unbounded number of available OAM states [5], mode division multiplexing at the terabit scale has been demonstrated [6].

Compared to the aforementioned conventional, *longitudinal* OAM, *transverse* OAM (t-OAM) of light was first theoretically predicted [7-9] and experimentally observed to be present in spatiotemporal optical vortices (STOVs) much more recently [10, 11]. The relation between the t-OAM and ST-TC of a STOV beam was also established recently [12]. STOVs carry t-OAM, which is characterized by a phase winding in the space-time plane, resulting in a phase singularity defined in the $x$-$t$ domain. Beams carrying t-OAM can now be generated using linear optical setups [11, 13], leading to burgeoning research interests, such as nonlinear optics of t-OAM. Notable examples include using second harmonic generation to perform *doubling* of ST-TC [14, 15] and using high harmonic generation to produce extreme ultraviolet STOV beams with high ST-TC [16]. The potential of STOV for communication was also demonstrated using STOV strings with 28 distinct charges [17]. These early successes underscore the promise of expanded t-OAM capabilities. To fully realize the potential of t-OAM beams for communication and information processing, it is necessary to develop arithmetic and logic operations that operate on this DOF. In recent years, it has been experimentally shown that arithmetic operations, which refers to addition, subtraction, multiplication, and division, can be performed on S-TC [18-24]. However, extending such arithmetic operations from the charges of conventional OAM to those of t-OAM is far from straightforward.

In this work, we demonstrate a novel optical computation pipeline capable of manipulating and transferring ST-TC. Specifically, we experimentally implemented, for the first time, the addition, subtraction, and readout of both integer and *fractional* ST-TC encoded in optical beams. As a proof-of-concept demonstration, our arithmetic operations manipulated spatiotemporal phases in the Fourier domain by utilizing pulse shapers with programmable spatial light modulators (SLMs). The sequential operations are performed using cascaded pulse shapers with proper relaying optics in between them. Mathematically speaking, our system applies a sequence of transformations (i.e., function composition) to the bosonic state defined by the ST-TC of the input light. Since operations on integer and fractional ST-TC are both feasible, our work demonstrates novel capabilities to perform arbitrary numerical addition and subtraction of ST-TC, and, hence, marks a crucial steppingstone toward full arithmetic operations on the ST-TC of light for bosonic state computation and information processing.

## 2 Information processing with spatiotemporal topological charge (ST-TC) of light

### *2.1 Overview of the ST-TC arithmetic pipeline via function composition*

In computing, an arithmetic logic unit (ALU) performs arithmetic and bitwise operations on integer binary numbers. They include addition, subtraction, increment, decrement, two's complement bitwise logical operations, bit shift operations, and other operations such as pass-through. Optical ALUs, including those using *longitudinal* OAM light and other photonics platforms [25, 26], have been demonstrated to be capable of performing such operations. Here, we demonstrate a new form of ALU that performs operations on the ST-TC of transverse OAM light. We extend the notion of bitwise operations to processes involving both integer and non-integer ST-TC values.

We begin by describing the optical computation pipeline and its experimental implementation using ST-TC of light. An overview of this method is schematically summarized in Figure 1, where a simplified data flow pipeline and its arithmetic operations is illustrated in Fig.1(a). The pipeline starts with a source with an initial ST-TC, $q_0$. This initial state is then sequentially transformed by three distinct processing modules, Device 1, Device 2, and Device 3, into the output states, $q_{out,1}, q_{out,2}, q_{out,3}$, respectively, before the readout of the final output ST-TC, $q_f$. Each module applies a specific operation, $f_1$, $g_1$, $f_2$, respectively. In this proof-of-concept two-operand demonstration, Devices 1 and 3 are a pair of cascaded pulse shapers. Each $n^{th}$ pulse shaper performs a summation of ST-TC sent into it (i.e., $q_{in}$) and the arbitrary charge $q_l$ associated with the Laguerre–Gaussian (LG) azimuthal mode index $l_n$ encoded on its phase mask. This is mathematically represented as $f_n(q_{in}, l_n) = q_{in} + q_l$. Device 2, placed between Devices 1 and 3, represents a relay system with unit magnification that ensures the imaging planes on each of the cascaded pulse shaper are appropriately mapped onto each other. Device 2’s operation is referred to as a pass-through in the language of an ALU, which can be expressed as $g_n(q_{in}) = q_{in}$. The function $g_{n=1}$ is used to represent the mathematical operation performed by such relay optics.

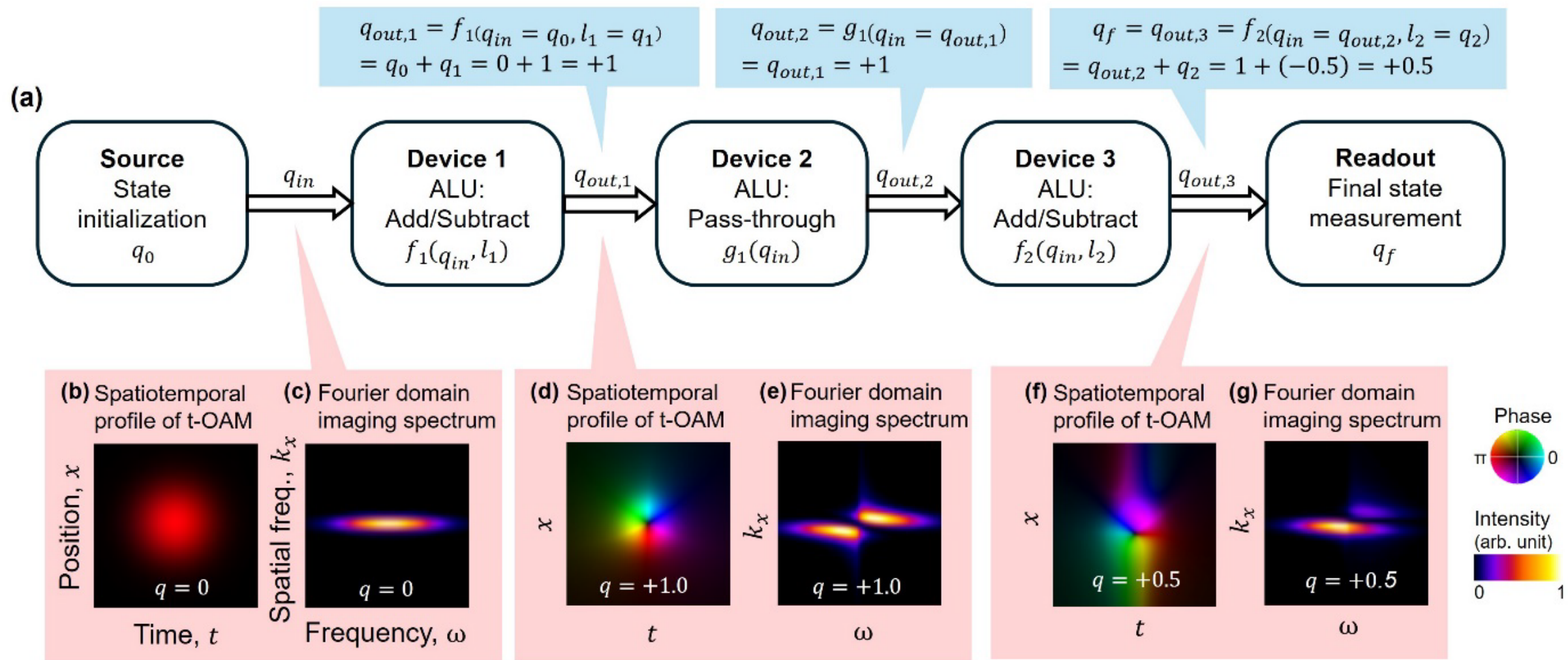

**Fig. 1** Optical computation pipeline implemented via ST-TC manipulation. This example demonstrates the addition and subtraction of two ST-TC values using three arithmetic logic units (ALUs). This architecture is scalable to accommodate a larger number of operands. **(a)** The information flow from a source, through the three devices, to the final readout. The blue shaded areas depict the ALU operations performed by each device as a mathematical function. Devices 1 and 3 perform the operation $f_n(q_{in}, l_n) = q_{in} + q_l$. The initial input state has ST-TC = 0 ($q_{in} = q_0 = 0$). In this example, Device 1 adds an ST-TC value of +1, and Device 3 adds an ST-TC value of -0.5 (equivalent to subtracting +0.5), resulting in a final state with ST-TC of $q_f$ = +0.5 at the readout. The function of the whole pipeline is $q_f = f_2 \circ g_1 \circ f_1(q_{in}) = f_2(g_1(f_1(q_{in}, l_1)), l_2)$. **(b-g)** The red shaded areas show the simulated corresponding ST-TC visualized in the space-time domain (x-t plane) (b, d, f) and Fourier domain ($k_x$-ω plane) (c, e, g) at each stage of the pipeline.

Next, the panel of Figs. 1(b) to (g) show the simulated ST-TC of light, in both space-time, *x-t* (b, d, f) and Fourier domain, $k_x$-$\omega$ (c, e, g), as it steps through the processing pipeline. Since the pulse shaper in our demonstration does not modify the light along the y-axis, this dimension is trivial and omitted from our discussion. The simulated complex field representations of the ST-TC of light in the *x-t* domain shown in Fig.1(b, d, f) follow the visualization scheme of Ref. [14, 27, 28], where brightness encodes light intensity, while hue encodes the spatiotemporal phase from 0 to 2π radians. Corresponding experimental $k_x$-ω results are presented in later sections. This panel of figures illustrates the case where an initial unstructured Gaussian light pulse without ST-TC, corresponding to a ST-TC of $q_0$ = 0, is used to perform a summation of ST-TC values $q_1 = 1$ and $q_2 = -0.5$. The resultant ST-TC of $q_f = 0.5$ is then directed to a readout system for decoding the final arithmetic result of the charge.

Turning to the inner workings of the system, we begin with Device 1 which serves as the first ALU. It performs the operation $q_{out,1} = f_1(q_{in}, l_1) = q_{in} + q_1$, where $q_1$ is the ST-TC to be added to the input. Experimentally, this operation can be realized using a non-cylindrically symmetric 4*f* pulse shaper, as demonstrated in previous works [29], or a specially designed single-pass optical element such as a slanted nanograting [30]. In our implementation, we adopt the former approach using a folded 4*f* configuration, in which a LG mode pattern with a digitally programmable azimuthal index *l* is applied to the SLM within the pulse shaper. When choosing $l_1 = +1$, Device 1 acts on the Gaussian pulsed beam ($q_0 = 0$) to produce the output state $q_{out,1} = f_1(q_{in} = q_0 = 0, l_1 = +1) = +1$. The simulated ideal spatiotemporal profiles of Device 1's output with ST-TC = +1 in *x*-*t* and $k_x$-$\omega$ domains are shown in Fig. 1(d) and (e), respectively. In these figures, defining features of ST-TC = +1 are clearly visible, namely a 2π phase winding in the *x*-*t* plane and a single gap separating two diagonally spaced lobes in the $k_x$-$\omega$ domain.

Device 2 is a second ALU that relays the input information to another location, which serves as a pass-through, analogous to an optical link or a transmission line. Its operating function is expressed as $q_{out,2} = g_1(q_{out,1}) = q_{out,1}$. Experimentally, Device 2 comprises an optical relay system based on a 4*f* imaging system with unit magnification. Its purpose is to transfer information encoded in a specific domain (e.g., *x*-*t* or $k_x$-$\omega$) from a designated space to another. Specifically, in the example given in Fig. 1, the 4*f* imaging system in Device 2 relays the $k_x$-$\omega$ plane that occurs at the SLM in Device 1 to the corresponding plane at the SLM in Device 3. Device 3 is a cascaded duplicate of Device 1 that serves a similar function. The difference is that it operates on the output of the preceding Device 2 and adds to it an ST-TC defined by the azimuthal mode index $l_2$ programmed on its SLM. In the example used in Fig 1, when choosing $l_2 = -0.5$, the output state then becomes $q_{out,3} = f_2(q_{in} = q_{out,2} = +1, l_2 = +0.5) = +0.5$.

The result of the arithmetic operations is ultimately directed to an optical readout system for decoding. Our pipeline is capable of handling *fractional* ST-TC values, distinguishing it from the majority of prior theoretical and experimental studies of STOVs. The spatial-temporal and its corresponding Fourier domain profiles of the resultant fractional ST-TC=+0.5 are illustrated in Fig.1(f) and (g), respectively. Since light carrying fractional ST-TC is more broadly distributed in space–time compared with that carrying integer SC–TC, the brightness of Fig. 1(f) was increased by a factor of 1.4× to improve visualization. In summary, the total operation of our proof-of-concept two-operand information-processing pipeline can be expressed as $q_f = f_2 \circ g_1 \circ f_1(q_{in}) = f_2(g_1(f_1(q_{in}, l_1)), l_2)$. We note that this approach is scalable to an arbitrary number of operands by concatenating additional optical ALUs, namely pulse shapers and relay systems.

### *2.2 Experimental implementation of the arithmetic pipeline in an optical platform*

As illustrated in Fig. 2(a), the experimental setup of a two-operand optical computation pipeline consists of five distinct modules. The first module, labeled Source in Fig. 2(a), is responsible for state initialization. It utilizes a Yb:KGW femtosecond laser oscillator (OneFive Origami 10-100) that delivers pulses with a central wavelength of 1028 nm, ~8 nm full-width at half-maximum (FWHM) spectral bandwidth, 200 mW average power, 100 MHz repetition rate, and a near transform-limited pulse duration of ~ 200 fs (FWHM). Following the laser oscillator, a Faraday isolator (Linos FL-1030-3SC), a 5× beam expander, and a power attenuator comprising a half-wave plate and a polarizing beam splitter are used to condition the beam for our state initialization.

Next in the pipeline are Devices 1 and 3. They are a pair of identical, cascaded pulse shapers. Each of the pulse shaper consists of a transmission optical grating (G1; 1600 lines/mm), a cylindrical lens (CL1; f = 300 mm), and a reflective spatial light modulator (SLM; Santec SLM-200) arranged in a 4*f* configuration. The SLMs can be programmed with phase patterns corresponding to LG azimuthal mode index *l*, enabling straightforward generation of arbitrary ST-TC, either integer or fractional, for a range of experimental tests. Device 2 is a 4*f* relay imaging system, comprising a pair of lenses, which maps the SLM plane of Device 1 onto that of Device 3 with a lateral magnification of -1. A phase correction on the SLM of Device 3 compensates for the inversion. The last module, labeled as Readout in Fig. 2(a), is an imaging spectrometer built in-house using a transmission grating (G3; 1739 lines/mm), a cylindrical lens (CL3; f = 130 mm), and a beam profiling camera (D; DataRay WinCamD). To enhance clarity, an unfolded schematic of our experimental setup following the Source is also shown in Fig 2(b).

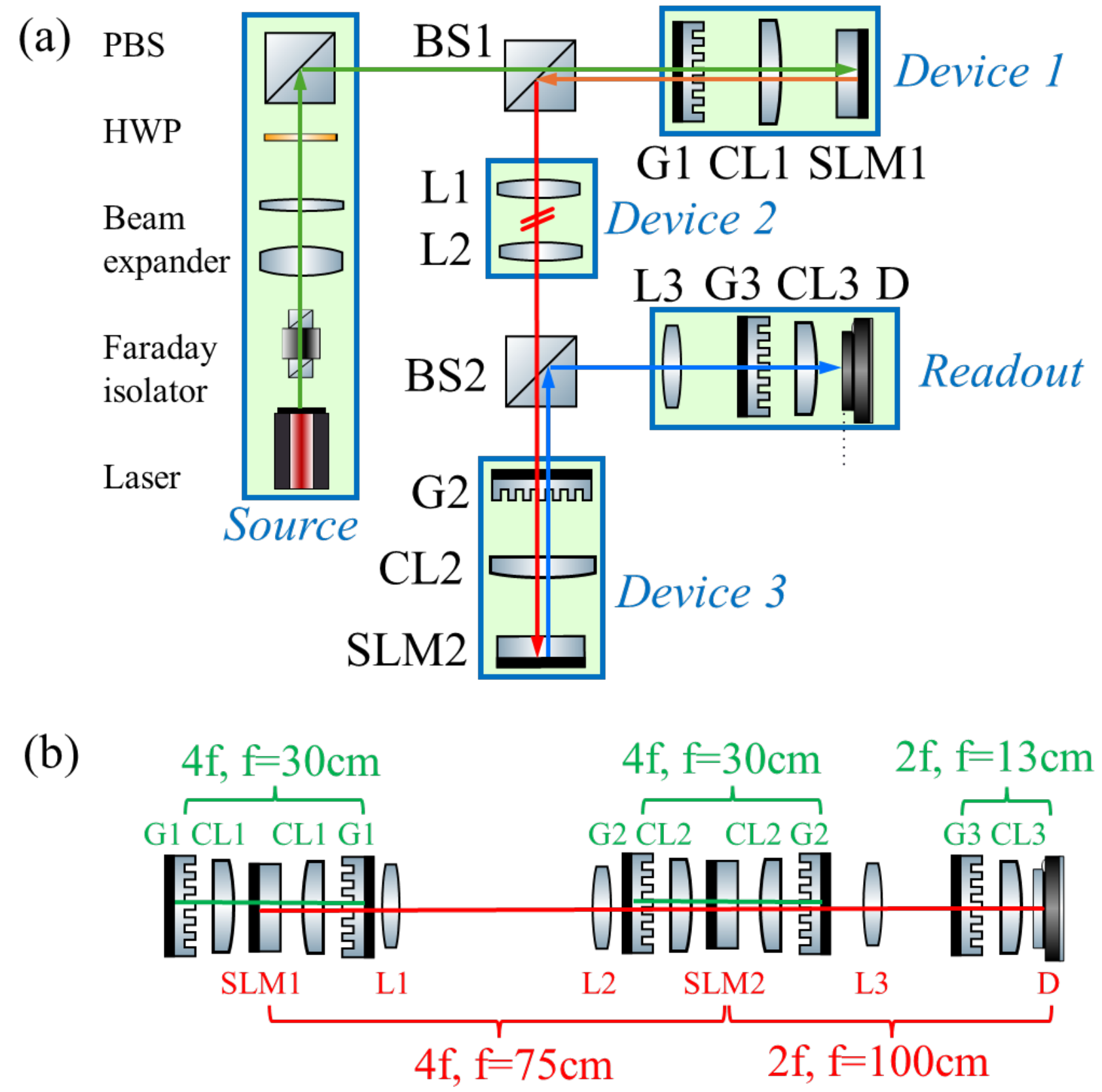

**Fig. 2** Experimental setup for arithmetic operations performed on the ST-TC of light. (a) The optical computation pipeline includes five modules (Source, Device 1, Device 2, Device 3, and Readout) as defined in Fig. 1. The Source is a femtosecond laser and associated components. Devices 1 and 3 are identical pulse shapers, Device 2 is a relay system, and the Readout device is an imaging spectrometer. (b) Unfolded schematic of the optical configuration. Each pulse shaper implements a 4*f* imaging configuration. The two SLMs and the relay system (SLM1, SLM2, L1 and L2) also form a 4*f* imaging system. Furthermore, SLM2, L3, and D constitutes a 2*f* imaging configuration. Additionally, G3, CL3, and D form another 2*f* imaging system. HWP: half wave plate; PBS and BS, polarizing and non-polarizing beam splitters; G, optical grating; CL, cylindrical lens; L, rotationally symmetric lens. SLM, spatial light modulator; D, 2D detector.

The readout imaging spectrometer, as demonstrated in Ref. [31], enables direct measurement of the resultant ST–TC value from the raw x–ω image. With suitable calibration of the x-axis, the x–ω image maps to $k_x$-ω, although this mapping does not alter the interpretation of ST–TC. We note that there are other alternative techniques that could characterize spatiotemporal structured light [32-34], including single-shot wavefront sensing techniques and ptychographic reconstruction methods [35, 36]. Our method measures the beam's $x$-$\omega$ profile, which is the Fourier-domain imaging spectrum of the SLM plane, using a simple imaging spectrometer without requiring interferometry with another reference beam [31]. The number of the gaps between bright lobes in the $x$-$\omega$ imaging spectrum corresponds to the absolute ST-TC value, $|q|$, while the orientation in which the lobes are aligned indicates the helicity of ST-TC (i.e., sign of q). In addition to the gaps between fully formed bright primary lobes, there are gaps between emergent dimmer secondary lobes that appear only with *fractional* ST-TC. Both our work and a recent preprint [37] have independently observed and demonstrated such distinguishing features, which will be discussed in greater detail in Section 3.

## 3 Experimental results

### *3.1 Fractional ST-TC and the importance of their initial azimuthal phase*

Given that research on fractional ST-TC is still in its nascent phase, we began by experimentally generating and characterizing these states using a single pulse shaper and employed the imaging spectrometer to identify their distinguishing features. Fig. 3(a) shows nine LG phase patterns imprinted on the SLM in the pulse shaper that produced the corresponding imaging spectra shown in Fig. 3(b). These LG patterns correspond to azimuthal mode indices $l$ = 0.5, 1, and 1.5, each implemented with *initial azimuthal phase angles* $\phi_0$ = 0°, 90°, and 180°, where $\phi_0$ is defined relative to the nine o'clock orientation on the SLM plane. Referring to the middle row of images in Fig. 3(a), which corresponds to $l$=1, we observe no azimuthal phase jump or discontinuity in the three LG patterns. This behavior holds for all integer values of $l$ and their resulting integer charges of $q$. In this context, *phase discontinuities* are defined as any abrupt phase jumps that are present after $2\pi l$ phase unwrapping, namely, modulo $2\pi$. In contrast, azimuthal phase discontinuities appear when fractional values of $l$ are used, leading to non-integer $q$. This feature is clearly visible in the top and bottom rows of Fig. 3(a). Moreover, the initial azimuthal phase angle, $\phi_0$, determines the position where the phase discontinuity occurs.

For instance, the discontinuities appear at the nine o'clock, twelve o'clock, and three o'clock positions when $\phi_0$ = 0°, 90°, and 180°, respectively.

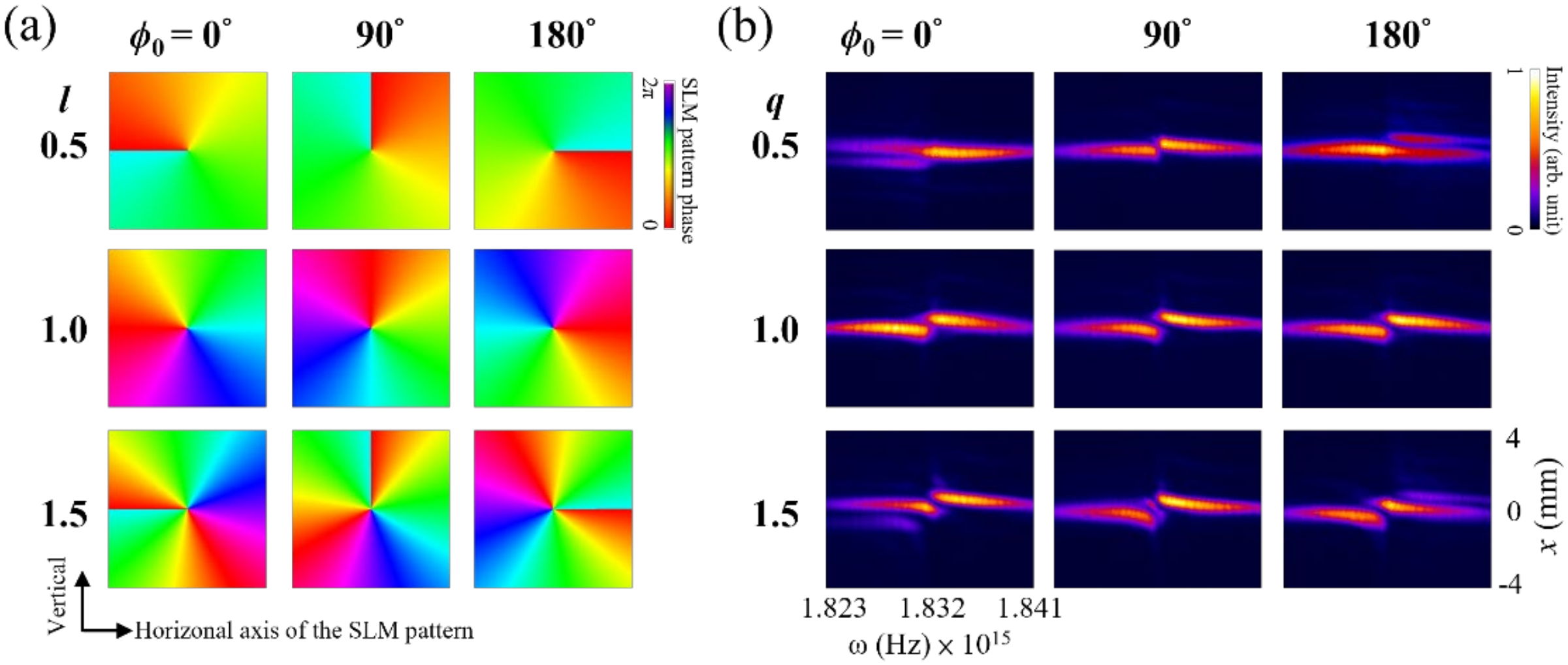

**Fig. 3** Experimental results illustrating the effects of initial phase conditions on fractional and integer ST-TC values. (a) SLM phase patterns in a pulse shaper with example LG azimuthal mode indices, *l (l* = 0.5, 1, and 1.5*)*, and initial azimuthal phase angles, $\phi_0$ ($\phi_0$ = 0°,90°, and 180°) are shown. Note that the azimuthal phase discontinuities are only visible on the top and bottom rows where the LG mode index is non-integer (*l*=0.5 and *l*=1.5). (b) Experimentally measured imaging spectra that correspond to the SLM patterns used in (a), leading to *q* = 0.5, 1, and 1.5, with different $\phi_0$. Observe that the broken lobe signature is easily observed on the top and bottom rows with fractional ST-TCs, except for the $\phi_0$=90∘ condition, where the broken lobe signature is obscured.

We experimentally demonstrate that the spectral imaging patterns corresponding to integer charges remain consistent as the azimuthal angle $\phi_0$ of the SLM patterns varies. Taking q=1 as an example, the middle-row images in Fig. 3(b) show that, regardless of the value of $\phi_0$, the x-ω images consistently exhibit two similarly aligned primary lobes. In contrast, in the case of fractional charges, the imaging spectra change with $\phi_0$. For fractional q, a weaker secondary lobe emerges adjacent to one of the bright primary lobes when $\phi$ = 0° or 180°. The brightness of this secondary lobe correlates with the difference between the fractional charge and the next higher integer charge (see Supplement Fig. S2 and Fig. S5). In addition to this unique feature, the location of the phase discontinuity on the phase mask determines the location of the emergent secondary lobe. For instance, when $\phi_0$ = 0°, the dislocation occurs on the left, i.e., the lower angular frequency side of the phase mask, and the secondary lobe appears *on the corresponding side* of the imaging spectrum. However, when ϕ = 90° or 270°, fractional ST-TCs are characterized by the appearance of smaller lobes nested between a pair of large lobes. For non-integer ST-TC values, the nested lobes are partially merged, and their degree of resolvability depends on the fractional component of the charge. As the fractional component approaches the next higher integer value, the nested lobes become increasingly well resolved (see Section 1 of the Supplementary Material). However, this present work focuses on ϕ = 0° and 180°, where the corresponding signatures can be resolved more readily and unambiguously. Furthermore, for all ϕ and ST-TC values, the lobes align along the diagonal for positive ST-TCs and along the anti-diagonal for negative ST-TCs (see Fig. S2, bottom right and top left, respectively).

These observations can be explained by the nature of the phase structure: while integer LG modes do not exhibit azimuthal phase discontinuity, fractional LG modes do, leading to the emergence of a secondary lobe in the *x*-*ω* plane, which can be directly observed on imaging spectra *only* when $\phi_0 \neq 90°$ or 270°. This is a new finding that was never reported in the literature, to the best of the authors' knowledge. A more detailed series of imaging spectra generated by varying *q* is provided in Supplementary Document (Supplement Section 2 and its corresponding Fig. S2 and Supplementary Media S1).

*3.2 Example demonstrations of the addition and subtraction of two ST-TCs*

Having established how an ALU operates when given an LG-mode pattern with an arbitrary azimuthal index *l*, we now cascade a second ALU to perform arithmetic operations between two ST-TCs, demonstrating the performance with both integer and fractional values. It is worth highlighting that a proper optical relay that images the Fourier plane of the first ALU onto the next is essential. Without it, STOV pulses become significant distorted during free-space propagation due to diffraction-induced phase effects [12, 38, 39]. Consequently, information encoding and processing along the pipeline are degraded, ultimately leading to failed arithmetic operations. We experimentally verified this, and the results are shown and discussed in the Supplementary Document (Supplement, Section 2 and its corresponding Fig. S2). These diffraction effects during free-space propagation of STOVs could potentially be mitigated by using optical fibers, an approach that is currently under intense investigation [40, 41].

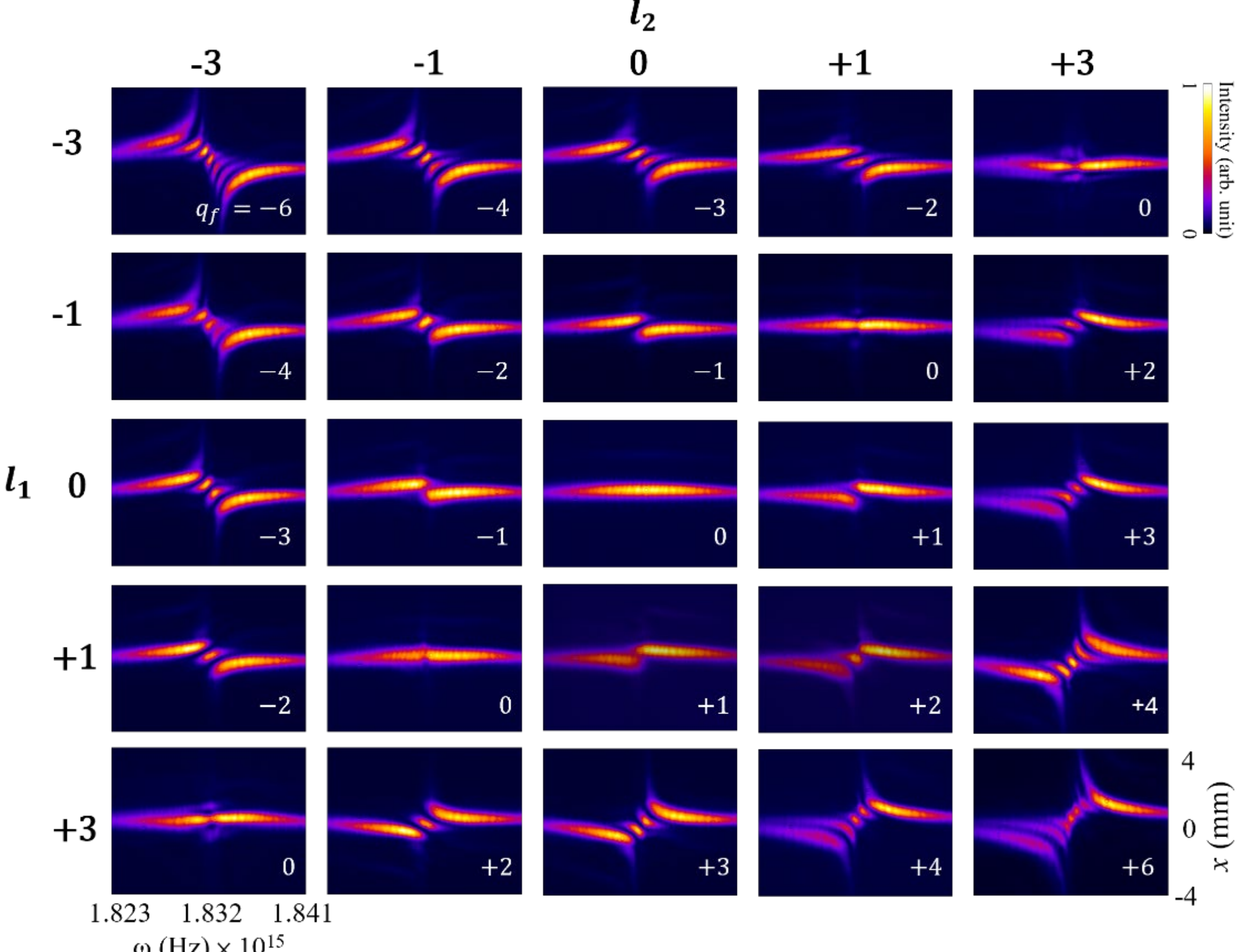

**Fig. 4.** Experimental results of a fully functional optical computation pipeline demonstrating the addition and subtraction of *integer* ST-TC values. The initial state has charge $q_0 = 0$. The first ALU (Device 1) is parameterized by $l_1 = 0, \pm1, \pm3$ (rows of the figure). The second ALU (Device 2) acts as a pass-through that does not change the ST-TC value. The third ALU (Device 3) is parameterized by $l_2 = 0, \pm1, \pm3$ (columns of the figure). After processing, the final readout exhibit charges $q_f = 0, \pm1, \pm2, \pm3, \pm4, \pm5, \pm6$, corresponding to the arithmetic operation $q_f = l_1 + l_2$.

Figures 4 and 5 present experimental results that validate the effectiveness and accuracy of our information-processing pipeline. Figure 4 reports the summation of various integer ST-TCs, while Figure 5 shows the corresponding fractional cases. Both use an initial state $q_0 = 0$. The values of the two operands are indicated in their respective rows and columns. The resultant ST-TC, $q_f$, of each optical computation pipeline is also labeled on each sub-figure. As before, integer $|q_f|$ can be decoded by counting the number of gaps between the bright primary lobes, while the sign of $q_f$ is inferred from their alignment along either the diagonal or anti-diagonal. For instance, in the top left of Fig. 4, seven lobes separated by 6 gaps and arranged anti-diagonally indicate $q_f = -6$. Conversely, in the bottom right of Fig. 4, a similar number of lobes are arranged diagonally, corresponding to $q_f = +6$.

A similar decoding process can be employed when fractional operands are involved, as experimentally showcased in Fig. 5. The key difference is that the number of gaps between the primary bright lobes determine the integer to the left of the decimal, while the relative brightness between the emergent secondary lobe and its adjacent primary lobe encodes the value to the right of the decimal. As such, all three factors—the number of gaps between primary lobes, the relative brightness between the secondary and adjacent primary lobe, and the orientation of the lobe pattern —must be considered together to fully decode the result. To enable efficient quantitative decoding, the relative brightness of the primary and secondary lobes for different fractional values can be calibrated and stored in a comprehensive lookup table, which is then used to infer fractional ST-TC values. In Supplementary Section 4 (Figs. S4 and S5), we present data for additional cases, each accompanied by line plots taken along the positive-slope diagonal and the negative-slope anti-diagonal. The plots demonstrate how brightness profiles would permit counting of primary peaks and quantification of secondary-peak intensities for ST-TC readout.

In summary, the representative readouts generated through our designed pipeline clearly showcase the accuracy and robustness of our optical arithmetic processes, demonstrating that it handles both integer and fractional operands equally well.

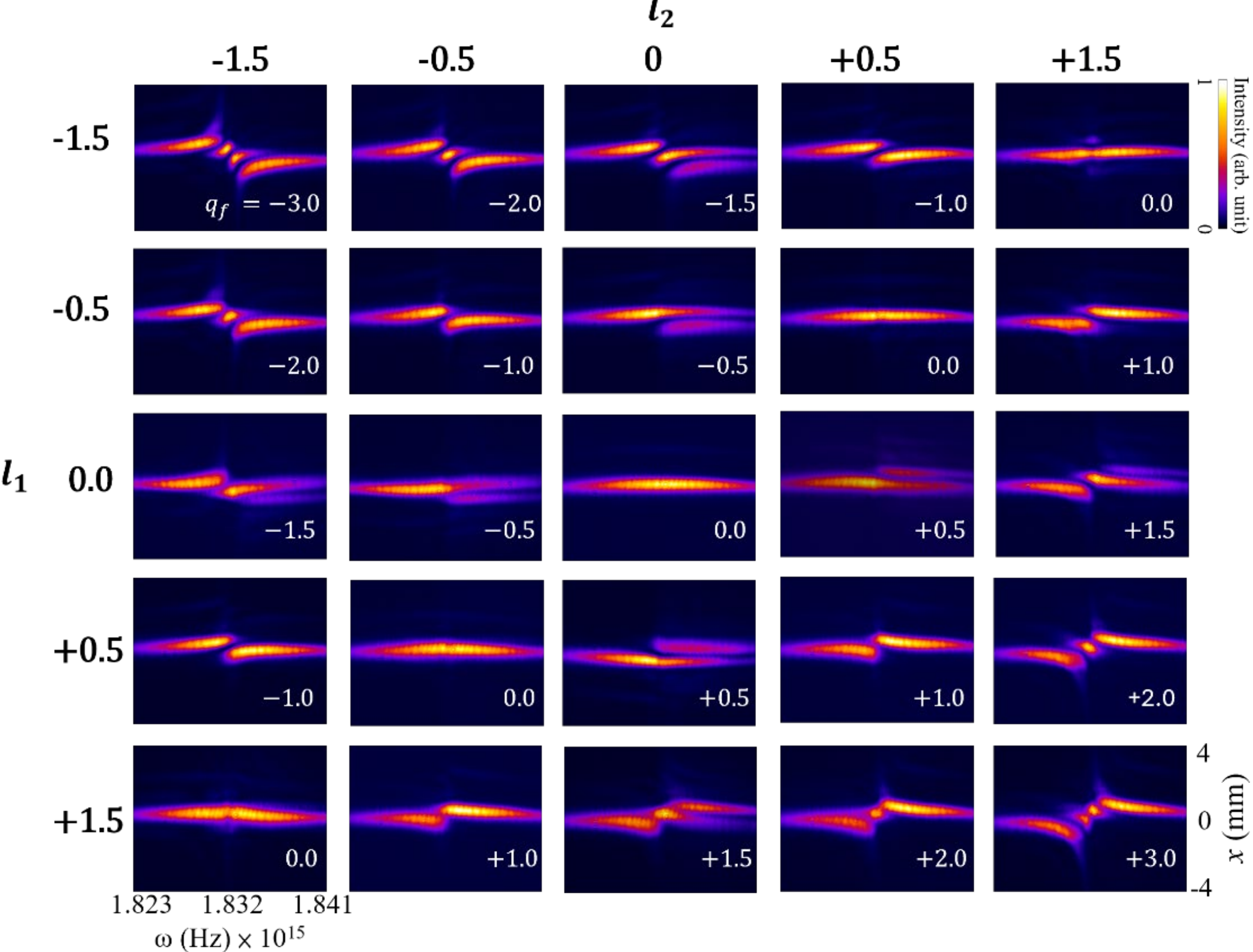

**Fig. 5** Experimental results of a fully functional optical computation pipeline demonstrating the addition and subtraction of fractional ST-TC values. The initial state has a charge $q_0 = 0$. The first ALU (Device 1) is parameterized by $l_1 = 0, \pm 0.5, \pm 1.5$ (rows of the figure). The second ALU (Device 2) acts as a pass-through that does not change the ST-TC value. The third ALU (Device 3) is parameterized by $l_2 = 0, \pm 0.5, \pm 1.5$ (columns of the figure). After processing, the final readout exhibit charges $q_f = 0, \pm 0.5, \pm 1, \pm 1.5, \pm 2, \pm 3$, corresponding to the arithmetic operation $q_f = l_1 + l_2$.

## 4 Discussion and Conclusion

A proof-of-concept two-operand ST-TC information-processing pipeline based on cascaded ALUs is presented and validated in this work. These ALUs perform arithmetic addition and subtraction operations, which can be represented as a function composition: $q_f = f_2 \circ g_1 \circ f_1 = f_2(g_1(f_1(q_{in}; l_1)); l_2)$. While a two-operand pipeline is presented here, we emphasize that this scheme can be readily extended and scaled to a longer pipeline, yielding $q_f = f_N \circ g_N \circ \ldots \circ g_4 \circ f_4 \circ g_3 \circ f_3 \circ g_2 \circ f_2 \circ g_1 \circ f_1(q_{in})$, where $N$ is an arbitrary number.

The primary limitations on N arise from the resolution of the decoding device and the cumulative optical loss that increases with *N*. This general pipeline is a mathematical function composition, which describes a sequential computation or information processing flow. An initial piece of data is sequentially transformed by *N* distinct processing modules, each applying a specific operation. The final transformed data is then sent to a readout module.

Furthermore, we note that the optical processes underlying the arithmetic operations in this work differ from approaches based on the in-phase and out-of-phase superposition of two optical beams carrying ST-TCs, as exemplified by Ref. [29]. In the present work, ST-TC manipulation is performed sequentially through cascaded arithmetic operations rather than through parallel manipulation followed by coherent combination. Specifically, the present approach implements a processing pipeline in which operations act *sequentially* on the optical mode. In contrast, the earlier approach relied on beam splitting, parallel manipulation of the resulting beams, and coherent recombination to yield a superposition of two ST-TCs.

In terms of precision, the arithmetic method presented in this work imposes, in principle, no lower bound on either the fractional operands it can process or the precision of the final arithmetic readout. In practice, however, the precision with which the final readout can be resolved is limited by the signal-to-noise ratio and the pixel density of the imaging spectrometer. The maximum decodable ST-TC is further constrained by experimental factors such as the SLM resolution, system noise, optical alignment, and diffraction. In the present implementation, our setup achieves a readout resolution of approximately $\Delta q \approx 0.1$.

Similar hardware-related and practical constraints also limit the maximum operand values that can be processed and the maximum achievable output ST-TC. Reaching higher ST-TC values requires higher-order LG modes with larger azimuthal indices, *l*. Under ideal Nyquist sampling, the present setup supports azimuthal indices up to $l \sim 1000$. However, while the SLMs in the ALU can generate large azimuthal indices for the operands, the detector used for the final x-ω spectral readout imposes additional practical constraints, similar to those that limit the readout precision. Under the current experimental conditions, high-quality x-ω spectral images were obtained for resultant ST-TC values up to $q_f$=30. Thus, within the limits of the present setup, operands with $|q|<1000$ can be processed, provided that the final output remains within $|q_f|<30$.

Additionally, while large amounts of group-delay dispersion (GDD) can affect STOV profiles, our setup contains no components that introduce substantial negative GDD, such as prism pairs, grating pairs, or chirped mirrors. The optical components in our system instead contribute an estimated positive GDD that is approximately two orders of magnitude below the level typically required to distort the STOV profile [42] and is therefore not expected to compromise decoding of the ST-TC from the measured profiles. Nevertheless, GDD in broadband pulses remains an important consideration in the formation of STOVs.

Beyond these experimental considerations, while the arithmetic demonstrated here is confined to simple addition and subtraction, the underlying principles are readily extensible to more sophisticated concepts in advanced disciplines. For instance, the pipeline described here is analogous to those commonly utilized in modern coherent optical communication systems, wherein information is encoded in the spatial, rather than spatiotemporal, phase of a light wave. In coherent optical communication systems, a highly stable single-frequency laser provides the optical carrier. For an initial unmodulated signal state (e.g., $q_{in}$=0) defined to be in phase with a reference beam, the carrier typically passes through an electro-optic phase modulator that adds or subtracts a phase $\pm l_1$ to the signal. The modulated signal then propagates through a transmission link before encountering another phase modulator or fixed-phase element that applies an additional $\pm l_2$. The resulting field is finally measured in a coherent receiver by interfering it with a reference laser, also known as the local oscillator, in an optical hybrid. A photodiode captures the interference pattern, representing the pipeline's final readout. Thus, the ALUs in our ST-TC processing pipeline perform roles analogous to each of the steps described above in a conventional coherent optical communication system. Of particular note, unlike conventional coherent detection, our ST-TC readout method is reference-free and requires only an imaging spectrometer, enabling direct inference of ST-TC values without computationally expensive interferometric reconstruction.

In addition to classical optical communication and computation, we anticipate that the same ST-TC processing scheme could be extended into quantum communication and computation, analogous to methods employed in continuous-variable quantum information science. For example, if a single-photon bosonic mode carrying t-OAM could be generated, it might encode a real number (e.g., ST-TC value) as an eigenvalue of its quadrature operator. An ALU, or a quantum logic gate, capable of performing addition or subtraction could then be realized using displacement operators acting upon the initial quantum state. In addition to arithmetic operations, pass-through modules could be incorporated, describing the temporal evolution of the quantum state of the photon's t-OAM during transmission between quantum gates. After several quantum logic gate operations and evolutions, the final readout, such as a homodyne projective measurement of the quadrature operator, would yield a classical measurement outcome, revealing the final ST-TC bosonic mode of the photon.

In conclusion, we have proposed and experimentally demonstrated a novel application of STOV beams carrying t-OAM of light. We first introduced both integer and non-integer charges of ST-TC in STOV beams and described our characterization method through imaging spectral analysis. We found that the initial azimuthal phase angle of LG mode patterns generated by a SLM within a pulse shaper is crucially important. Specifically, this initial phase angle uniquely determines the Fourier-domain imaging spectral patterns only for fractional but not integer ST-TCs, which is a previously unreported and distinctive signature. Subsequently, we proposed and constructed an optical computation pipeline leveraging the ST-TC of light. As a proof-of-concept demonstration, we built a two-operand arithmetic pipeline based on four ALUs: two modules designed to perform arithmetic operations on ST-TCs, one pass-through module that facilitates information transmission via optical relay, and a readout device. To demonstrate basic arithmetic capabilities, we successfully conducted addition and subtraction operations on both integer and non-integer ST-TC values using cascaded modules. Our comprehensive experimental results validate the accuracy of these arithmetic operations and confirm the practical applicability of ST-TC based optical information processing. The proposed processing pipeline presented here is highly versatile, with the underlying principles directly extendable to various applications in both classical and quantum optical computing and communication. We anticipate that optical

ST-TC will enable numerous future innovations across diverse disciplines, including communication, computing, imaging, cryptography, and beyond.

### *Funding*

This research is supported by the U.S. Department of Energy (DOE) Office of Science, Office of Biological and Environmental Research (BER), grant no. DE-SC0025194.

### *Acknowledgment*

The authors acknowledge Kefu Mu from the Department of Intelligent Systems Engineering, Indiana University Bloomington, for experimental assistance and discussion.

### *Disclosures*

The authors have a relevant provisional patent application related to this work.

### *Data availability*

Data underlying the results presented in this paper are not publicly available at this time but may be obtained from the authors upon reasonable request.

# Arithmetic with spatiotemporal optical vortex of integer and fractional topological charges: Supplemental material

## 1. Characterization of STOVs at an initially azimuthal phase angle of $\phi_0$ = 90°

This study focuses on operating the optical computation pipeline at initial azimuthal phase angle $\phi_0$ = 0°, which applies equally to the case $\phi_0 = 180°$, since the features used to decode q from the x–ω graphs are identical except for a $180°$rotation. In contrast, the decoding method must be modified at $\phi_0$ = 90° since the characteristic features differ. Without loss of generality, in this section we present simulated x-ω profiles for only $\phi_0 = 90°$, since the corresponding features at $\phi_0 = 270°$are related by a trivial $180°$rotation, in the same manner as the features at $\phi_0 = 180°$are related to those at $\phi_0 = 0°$.

From the simulated images shown in Fig. S1, it is shown that at $\phi = 90°$, the values of $q$ can be extracted by analyzing the number of smaller lobes nested between a pair of larger lobes and their degree of separation. At integer values, there are $(q - 1)$ well-resolved nested lobes, with well-defined gaps between adjacent lobes. However, when q is fractional, the nested lobes are not as well resolved. As q increases from a negative to a positive value in fractional steps, the evolution of the x-ω profile is marked by the emergence of additional nested lobes that are initially connected to each other but become increasingly separated as the fractional value approaches the next higher integer charge.

Based on these results, the optical computation pipeline described in the manuscript for ϕ = 0° and 180° can be directly extended to the cases of ϕ = 90° and 270°. Specifically, the modulation contrast and peak-to-peak distance associated with the charge-dependent nested lobes can be extracted from the diagonal and anti-diagonal lineout plots and then calibrated as functions of the ST-TC value, enabling construction of the corresponding lookup table. Although this lookup table would differ from that for ϕ = 0° and 180°, the overall methodology remains unchanged.

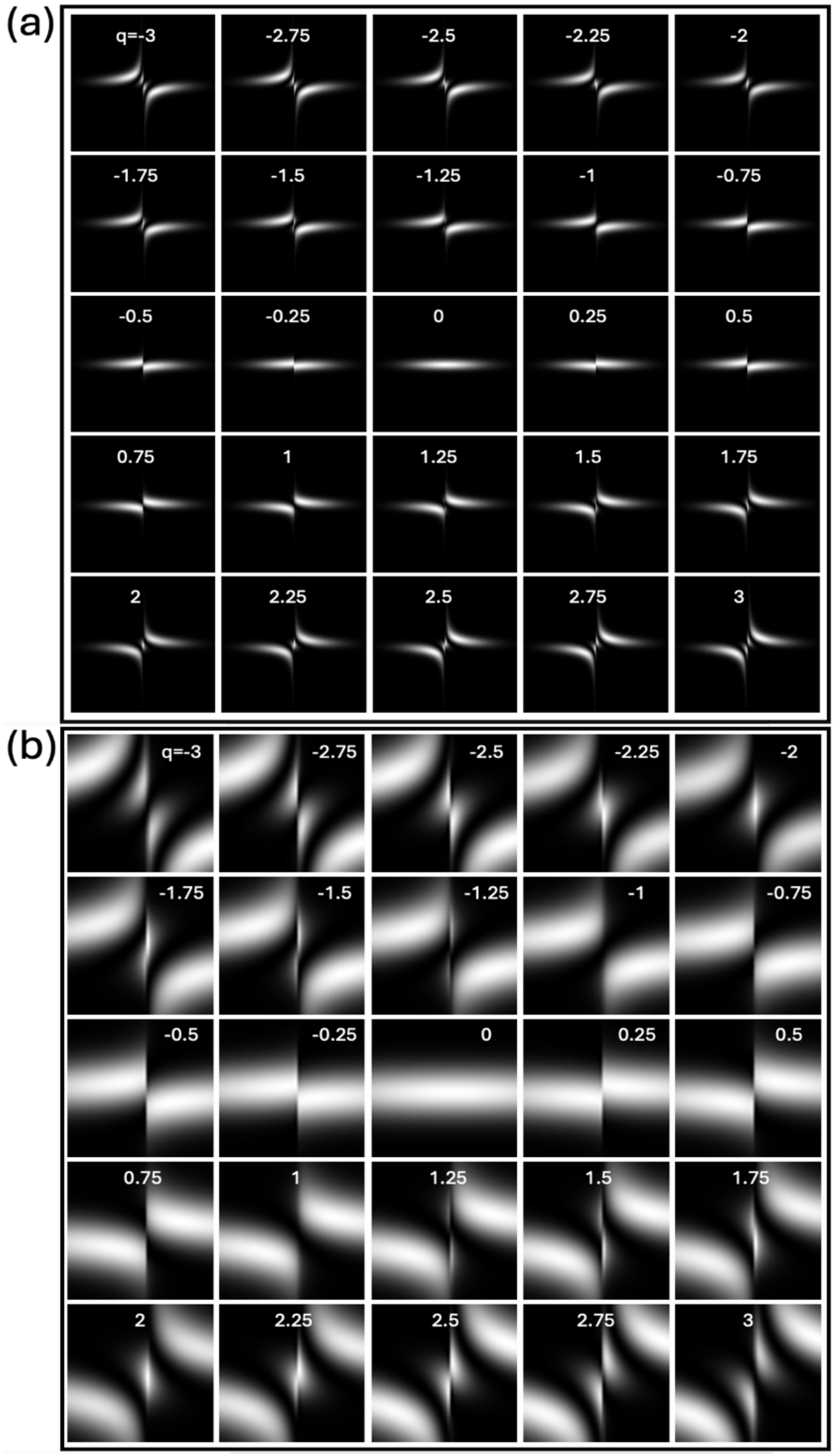

Fig. S1. Simulated x-ω profiles at $\phi_0$ = 90°. (a) x-ω profiles for $q_f$ values ranging from -3 to 3 in increments of 0.25. (b) Enlarged views of the profiles, illustrating the evolution of the nested lobes over the same $q_f$ range.

## 2. Evolution of imaging spectra of fractional and integer spatiotemporal topological charges (ST-TCs) of light

Experimentally measured imaging spectra corresponding to ST-TC values ranging from $q = -2$ to 2, in steps of 0.2, are shown in Fig. S2. Here, we clearly observe the emergence of secondary lobes, whose brightness increases as the fractional charge approaches the next higher integer value, as discussed in the main manuscript. Exploitation of this feature for decoding purposes is discussed in Section 3 of this supplemental document. A series of imaging spectra is compiled into a supplemental video, Supplementary Media S1 (Visualization 1), to facilitate visualization of the gradual changes exhibited.

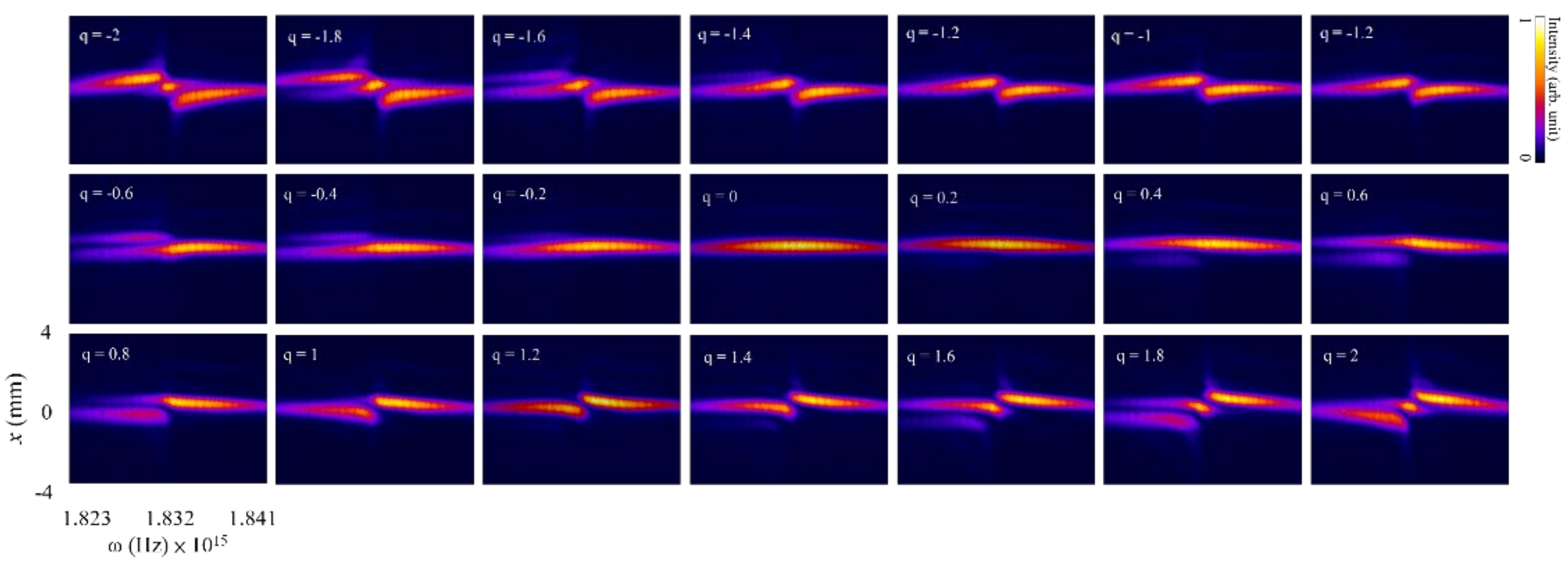

Fig. S2. Experimentally measured imaging spectra corresponding to ST-TC values ranging from $q$=−2 to $q$=2 in increments of 0.2.

## 3. Demonstrating the importance of a 4*f* pass-through relay system

Precise placement of the 4*f* relay optics is essential for accurately mapping the SLM plane of one pulse shaper onto that of the next, thereby ensuring high performance of the optical arithmetic pipeline. Any deviation from an ideal 4*f* optical relay system would lead to a blurred imaging spectrometer output, which prevents accurate decoding of the ST–TC. This is demonstrated by transmitting ST–TC values (i.e., $q$) between two cascaded pulse shapers through a pair of lenses that were intentionally configured to deviate from the 4*f* configuration, as shown in Fig. S3(a). This pass-through module is referred to as Device 2 in the main manuscript and forms part of the arithmetic pipeline. Fig S3(b) presents five experimental imaging spectra obtained from this aberrant system, with $l_2 = 0$ and $l_1 = -3$, −1, 0, 1, 3. It is readily apparent that a misaligned 4*f* system results in substantially degraded imaging spectra, which impairs our ability to resolve the number of lobes required for ST-TC decoding. For instance, in Fig. S3(b), the gaps between lobes that enable accurate decoding are obscured as the lobes become distorted and merge.

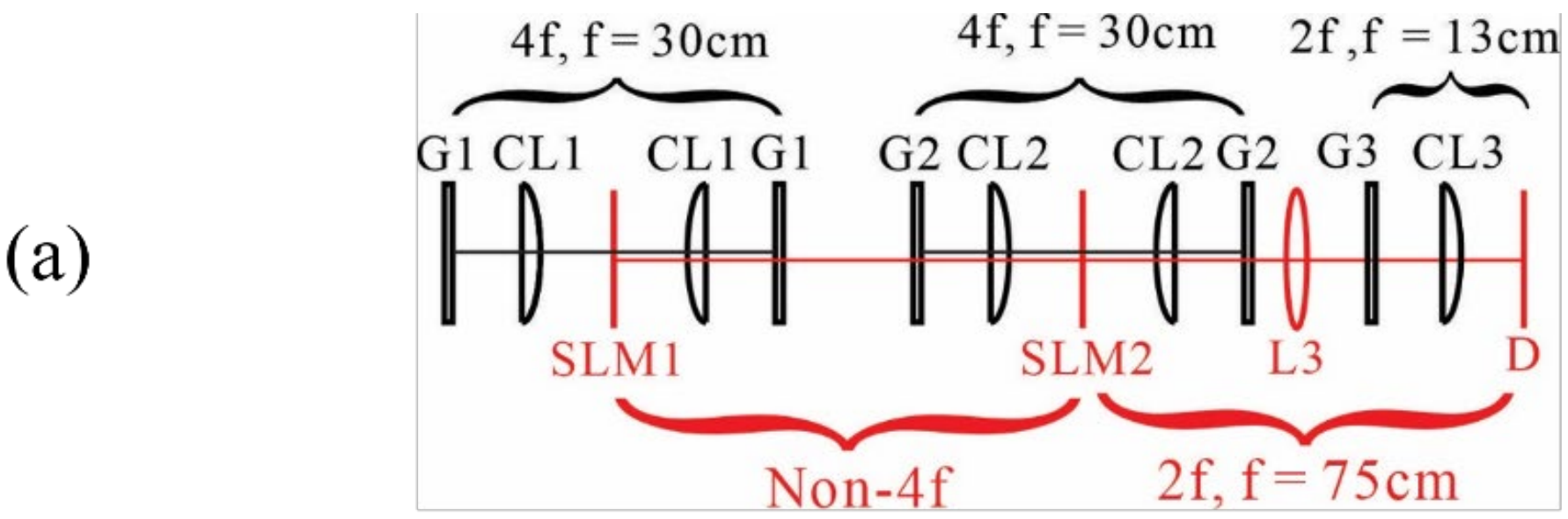

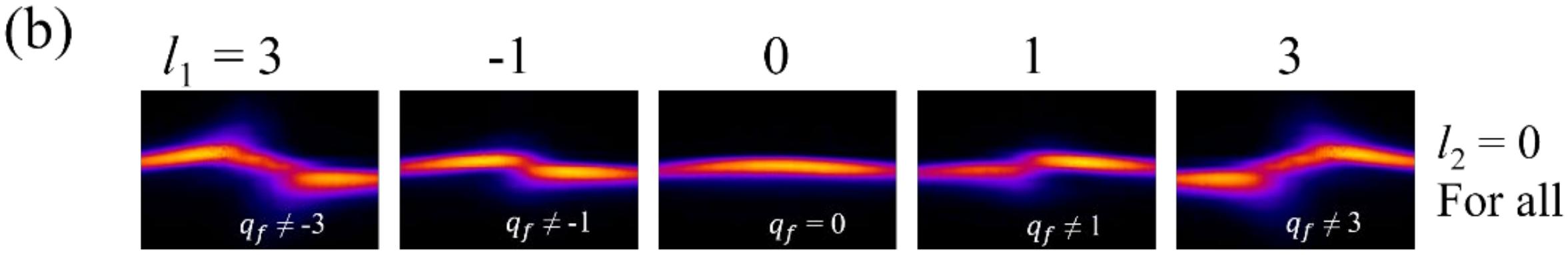

Fig. S3. Experimental demonstration of the importance of accurate 4f imaging between SLM planes in cascaded pulse shapers. (a) Schematic illustrating the intentional disruption of the 4*f* imaging configuration. (b) The resultant aberrant system exhibits severely blurred x-ω profiles, which prevents reliable evaluation of the arithmetic operation results.

## 4. Quantifying fractional and integer ST-TCs from imaging spectral analysis

Quantitatively decoding of integer ST–TC values can directly be achieved by determining the number of gaps between the bright primary lobes present in the raw x–ω spectral images. In contrast, the x-ω profiles of fractional ST-TC values exhibit bright primary lobes accompanied by weaker secondary lobes, whose relative brightness increases with the fractional component of ST-TC. In this section, we demonstrate that, instead of the full x-ω profile, the decoding process can be simplified by using one-dimensional diagonal (positive slope) and anti-diagonal (negative slope) lineouts (i.e., intensity curves along the diagonal or anti-diagonal of the x-ω profiles), regardless of whether the ST–TC values are integers or fractional.

We first consider the case when the value of ST-TC (i.e., $q$) is an integer value. By experimentally varying the LG mode indices $l_1$ and $l_2$ encoded on the SLM within each cascaded pulse shaper in the arithmetic pipeline, and by measuring the x-ω profiles of the results, a set of 25 different resultant x-ω profiles, along with their corresponding diagonal (blue) and anti-diagonal (red) lineouts, are shown in Fig. S4. When the arithmetic operations result in $q_f = 0$, such as when ($l_1$, $l_2$) = (–3, 3), (–1, 1), (0, 0), (1, –1), and (3, –3), there will only be a single horizontal lobe present, and both the red and blue lineouts produce a single peak. In contrast, when $q$ is a positive or negative integer value, multiple lobes appear along either the diagonal (blue lines in Fig. S4) or anti-diagonal (red lines) axis, respectively. Consequently, the lineouts along the corresponding directions exhibit high contrast modulation, with the number of dips equal to $|q_f|$. Therefore, depending on the sign of $q_f$, only either the anti-diagonal or diagonal lineout is needed for decoding the resultant ST-TC value. The use of the lineout along the other axis is unnecessary.

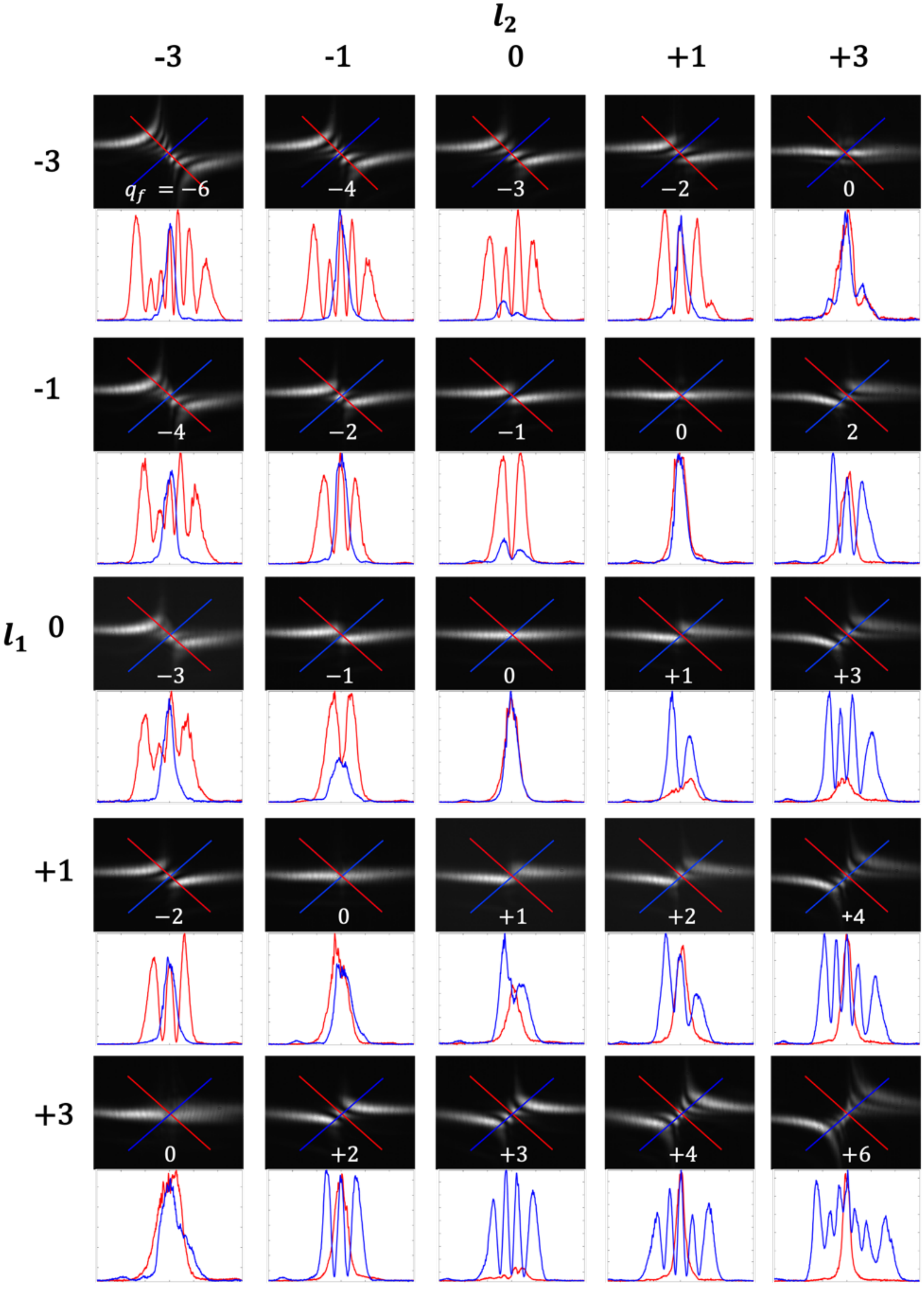

Fig. S4. Experimentally measured x-ω profiles and their corresponding diagonal (blue; positive slope) and anti-diagonal (red; negative slope) lineouts for different resultant $q$ values arising from different combinations of integer LG mode indices, $l_1$ and $l_2$.

We next consider fractional ST-TC. We simulate the x–ω profiles for q values (i.e., $q_f$) from −3 to 3 with a step size of 0.25. The x-ω profiles, together with the corresponding diagonal (blue) and anti-diagonal (red) lineouts, are shown in Fig. S5. Regardless of whether $q_f$ is an integer or a fractional value, the lobes are distributed along the diagonal for positive $q_f$ and along the anti-diagonal for negative $q_f$, consistent with the integer $q_f$ case. The main difference is that, for fractional values, the x-ω profiles contain a number of bright primary lobes equal to the integer part of $q_f$, along with an additional, dimmer secondary lobe. Furthermore, the relative brightness of the secondary lobe correlates to the fractional component of $q_f$. Specifically, as $q_f$ approaches the next integer, the secondary lobe becomes brighter, and this trend is captured by the lineouts and can be used for ST–TC decoding. Lineouts along the lobe direction are symmetric for integer $q_f$. For fractional $q_f$, however, they become asymmetric and exhibit an overall intensity shift relative to the center that grows with the fractional part. This behavior results in a slight offset between the central maxima or minima of the diagonal and anti-diagonal lineouts that varies with the fractional part of $q_f$. Additionally, the secondary-to-primary peak ratio reflects the magnitude of the fractional part. This symmetry-based feature remains robust even when the lineout modulation associated with the secondary lobe is extremely weak (for example, see $q_f$ = 2.25 in Fig. S5). Moreover, for odd $q_f$ the center of the x-ω profile falls in the gap between adjacent lobes, whereas for even $q_f$ it coincides with the central lobe. When the lobe pattern transitions between center-bright and center-dark as $q_f$ varies fractionally, a lineout taken perpendicular to the lobe alignment captures the corresponding change in intensity.

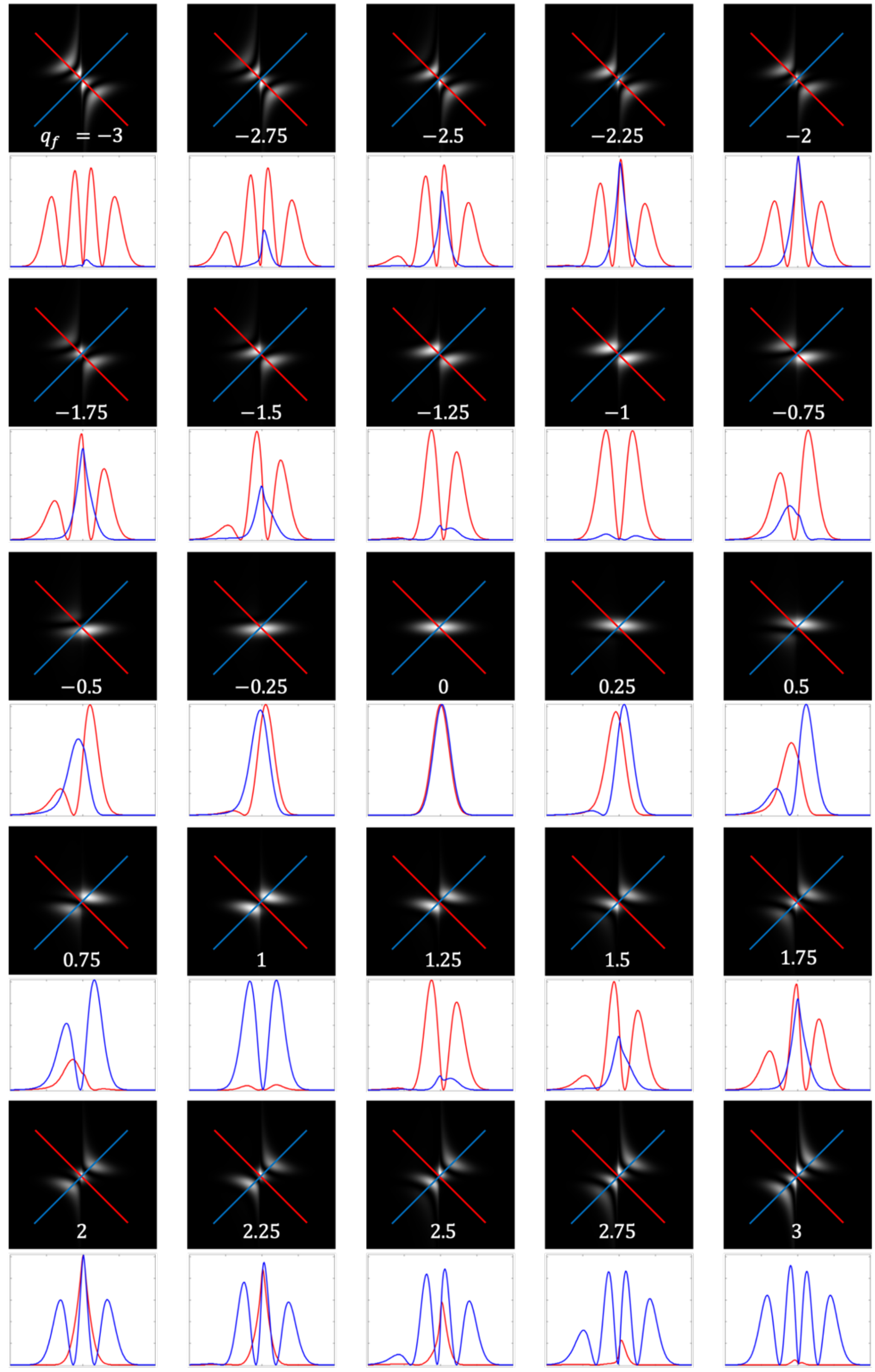

Fig. S5. Simulated x-ω profiles and their corresponding diagonal (blue; positive slope) and anti-diagonal (red; negative slope) lineouts for various $q_f$ values ranging from −3 to 3 in increments of 0.25.

We further verified these features experimentally by varying the LG mode indices, $l_1$ and $l_2$, encoded on the SLM in each cascaded pulse shaper within the arithmetic pipeline, and by measuring the resulting x-ω profiles and corresponding lineouts. Here, $l_1$ and $l_2$ were varied from -1.5 to +1.5 in steps of 0.5. The resulting 25 x-ω profiles, together with their corresponding diagonal (blue) and anti-diagonal (red) lineouts, are shown in Fig. S6. These results show that the features required to decode both integer and fractional ST–TC are largely observed experimentally, confirming the feasibility of the decoding strategies described above. While minor differences relative to the simulations are present, they are likely attributable to lens aberrations and optical alignment that slightly modify the relative brightness of the lobes. Such effects could be mitigated with improved optical engineering, which is beyond the scope of this paper, and any residual experimental variations can be accounted for through calibration. Because the aspect ratio of the measured x-ω profiles depends on the experimental configuration, the lineout angles must be adjusted accordingly. Overall, our results indicate that the relative peak prominence in the lineouts can be calibrated as a function of $q_f$ and stored in a comprehensive lookup table, enabling rapid decoding of arbitrary ST–TC values.

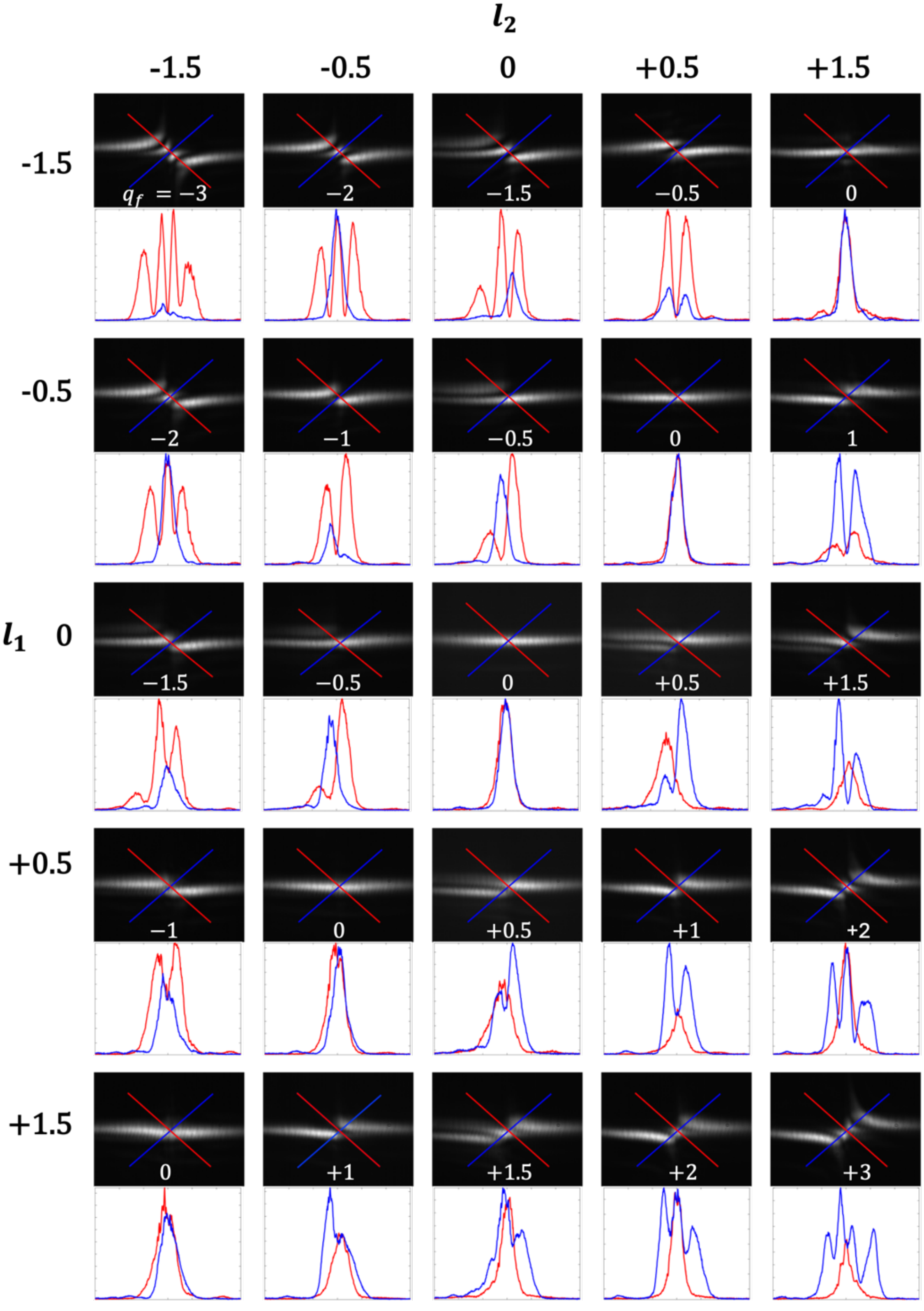

Fig. S6. Experimentally measured x-ω profiles and their corresponding diagonal (blue) and anti-diagonal (red) lineouts for different combinations of integer and fractional LG mode indices, $l_1$ and $l_2$.